\documentclass[lettersize,journal]{IEEEtran}
\usepackage{amsmath,amsfonts}
\usepackage{algorithmic}
\usepackage{algorithm}
\usepackage{array}
\usepackage[caption=false,font=footnotesize]{subfig}
\usepackage{textcomp}
\usepackage{stfloats}
\usepackage{url}
\usepackage{verbatim}
\usepackage{graphicx}
\usepackage{cite}

\usepackage{makecell}
\usepackage{xcolor}
\usepackage{multicol}
\usepackage{enumitem}
\usepackage{multirow}
\usepackage[capitalise]{cleveref}

\begin{document}

\title{An Empirical Study on Fault Detection and Root Cause Analysis of Indium Tin Oxide Electrodes by Processing S-parameter Patterns}

\author{Tae Yeob Kang,~\IEEEmembership{Member,~IEEE,}
        Haebom Lee,
        Sungho~Suh,~\IEEEmembership{Member,~IEEE}
		\IEEEcompsocitemizethanks{\IEEEcompsocthanksitem 
          This study was supported by a research grant from The University of Suwon, Republic of Korea in 2023. \protect 
          \IEEEcompsocthanksitem
            T. Y. Kang is the corresponding author and is with Department of Mechanical Engineering, The University of Suwon, Hwaseong, 18323, Korea (e-mail: tykang@suwon.ac.kr). 
            H. Lee is with Department of Computer Science, Heidelberg University, 69117 Heidelberg.
            S. Suh is with Embedded Intelligence, German Research Center for Artificial Intelligence (DFKI), 67663 Kaiserslautern, Germany.  
            \protect
		}
}

 \markboth{Preprint}%
{Kang \MakeLowercase{\textit{et al.}}: An Empirical Study on Fault Detection and Root Cause Analysis of Indium Tin Oxide Electrodes by Processing S-parameter Patterns}


\maketitle

\begin{abstract}
In the field of optoelectronics, indium tin oxide (ITO) electrodes play a crucial role in various applications, such as displays, sensors, and solar cells. Effective fault diagnosis and root cause analysis of the ITO electrodes are essential to ensure the performance and reliability of the devices. However, traditional visual inspection is challenging with transparent ITO electrodes, and existing fault diagnosis methods have limitations in determining the root causes of the defects, often requiring destructive evaluations and secondary material characterization techniques. In this study, a fault diagnosis method with root cause analysis is proposed using scattering parameter (S-parameter) patterns, offering early detection, high diagnostic accuracy, and noise robustness. A comprehensive S-parameter pattern database is obtained according to various defect states of the ITO electrodes. 
Deep learning (DL) approaches, including multilayer perceptron (MLP), convolutional neural network (CNN), and transformer, are then used to simultaneously analyze the cause and severity of defects. Notably, it is demonstrated that the diagnostic performance under additive noise levels can be significantly enhanced by combining different channels of the S-parameters as input to the learning algorithms, as confirmed through the t-distributed stochastic neighbor embedding (t-SNE) dimension reduction visualization of the S-parameter patterns.

\end{abstract}

\begin{IEEEkeywords}
Deep learning, Fault diagnosis, ITO transparent electrodes, Root cause analysis, S-parameter 
\end{IEEEkeywords}

\section{Introduction} 
        
\begin{figure}[!t]
		\begin{center}
			\includegraphics[width=\columnwidth]{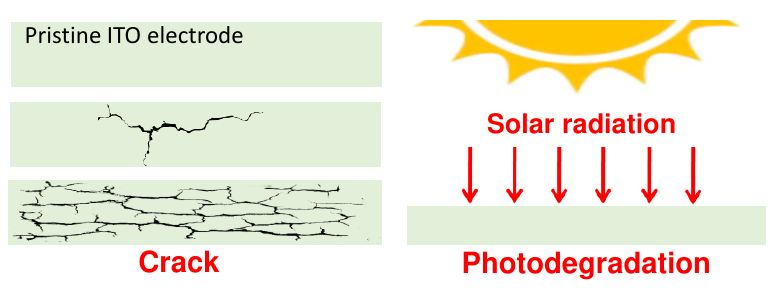}

   \vspace{-4mm}
			\caption{Major defects in ITO electrodes} 
			\label{fig:overview}
		\end{center}
       \vspace{-9mm}
	\end{figure}

    \IEEEPARstart{I}{ndium} tin oxide (ITO) electrodes are the basic building blocks of modern optoelectronic devices, such as photovoltaic cells, liquid crystal displays, organic light-emitting diodes, and various sensors, due to their high electrical conductivity, high optical transparency, and good physical and chemical properties \cite{ITOreview1_1,ITOreview1_2}. However, harsh operational environments induce defects in the electrodes and degrade their intrinsic properties, affecting the overall device performance. \cref{fig:overview} shows the major types of defects in ITO electrodes and their root causes. For instance, the ITO layer can crack when ITO-coated substrates undergo mechanical stress, such as bending or stretching. This is because the ITO film is usually deposited on substrates, such as polymer or glass, which can deform under stress. As a result, the ITO film suffers from tensile or compressive stresses that can lead to microscale and macroscale cracks \cite{ITO_macrocrack, ITO_microscale}. Moreover, 
    in numerous applications, ITO thin films are likely to be exposed to solar radiation which leads to photodegradation over time\cite{ITO_photo,ITO_chemical}.         
    Meanwhile, it is critical to ensure the reliability of the ITO electrodes for the overall health and performance of the device. Timely fault diagnosis and root cause analysis allow for preventive maintenance and effective rework. So far, numerous approaches, such as image processing, scanning probe, and non-scanning probe techniques, have been explored for defect inspections of ITO electrodes and related industrial products \cite{ITOreview1}. 

\begin{figure}[!t]
		\begin{center}
			\includegraphics[width=\columnwidth]{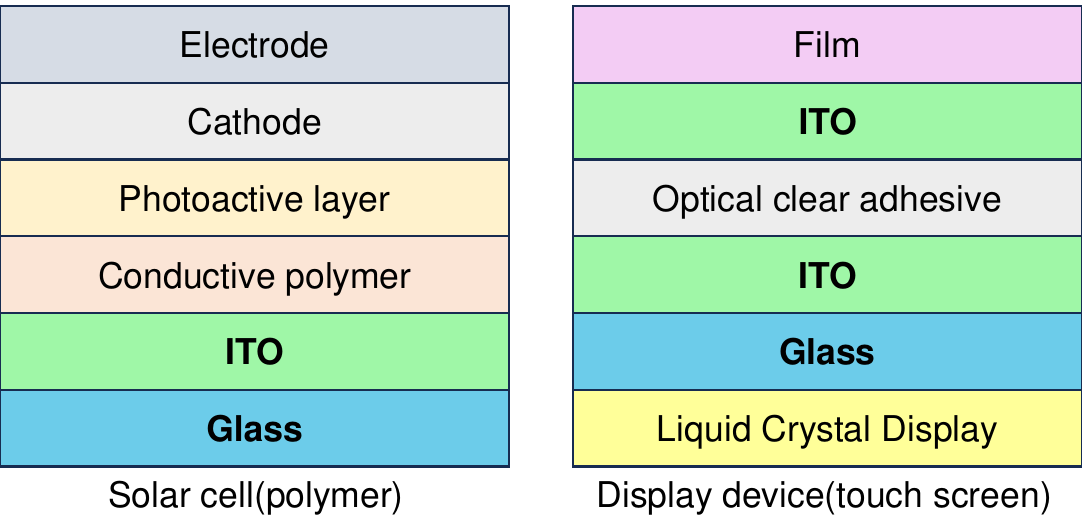}
			\caption{Simplified structures of optoelectronic devices including ITO electrodes: the ITO electrodes are placed between other layers} 
			\label{fig:ITO_structure}
		\end{center}
        \vspace{-8mm}
	\end{figure}

Due to its efficiency, image processing has been widely employed in manufacturing sectors to improve visual inspection for optoelectronic devices \cite{ai2002analysis, agrawal2011glass}. The image processing techniques have been utilized to detect, identify, and classify various types of defects by extracting optical features \cite{izmirliouglu2015glass,akdemir2015glass}. However, it is difficult to detect and diagnose onset defects in the transparent ITO electrodes. Hence, to assist the image processing, non-probing scanning techniques, such as electron microscopy \cite{ITO_review_8}, terahertz-time domain spectroscopy \cite{ITO_review_10}, and thermography \cite{ITO_review_11} have been developed, providing an excellent measurement throughput. Furthermore, an automated defect detection and localization method in photovoltaic cells using semantic segmentation of electroluminescence images has been proposed \cite{aziz2020novel}. However, their practical applications are limited. For instance, electron microscopy has such a small field of view; thermography requires electrical contact; and terahertz-time domain spectroscopy requires nonconventional, expensive, and complicated measurement systems. For further investigation of the defects in the ITO electrodes, conventional methods based on scanning probes can be utilized. These methods include the four-point probe method \cite{ITO_review_3, ITO_review_4}, conducting tip atomic microscopy \cite{ITO_review_5}, eddy current probe microscopy \cite{ITO_review_6}, and scanning probe microwave microscopy \cite{ITO_review_7}, which are applied in various types of research on electrical properties of thin films. However, due to the extremely low measurement throughput of scanning methods, they cannot be used for practical applications that demand a high measurement throughput. Furthermore, these methods target bare ITO electrodes. Therefore, it would be difficult to apply the aforementioned methods to the ITO electrodes manufactured in the optoelectronic devices as shown in \cref{fig:ITO_structure}. In reality, the electrodes are placed between other functional layers, eventually requiring the disassembly of the structures.

To avoid the disassembly of the packages and provide high measurement throughput, electrical signals between the ends of the electrode have been utilized for in situ fault diagnosis of electrodes in electronic packages. The related research works are as follows. The DC resistance measurement method was widely used for reliability monitoring of electrodes because of its simplicity and convenience. DC resistance accurately responds to a short or an open-state conductor. However, it is not suitable to indicate the evolution of defects \cite{kwon2010detection}. To overcome the limitation, studies have suggested electronic parameters with high-frequency components that take advantage of the skin effect for early detection of defects. For instance, Kwon et al. \cite{kwon2009early, kwon2015remaining} demonstrated the value of radio frequency (RF) impedance measurements as an early indicator of the physical degradation of solder joints as compared to DC-resistance measurements. RF impedance is sensitive to defects initiated on the surface due to the skin effect, proving the feasibility of detecting defects in the electronic components. Scattering parameters (S-parameters) can be used to describe the general behavior of electrodes \cite{bogatin2004signal}; hence they can also be utilized as an indicator of defects. Kruger et al. \cite{kruger2009measurement} showed that the sensitivity of the S-parameters was better than that of the DC resistance as the crack propagated in the electrode. Putaala et al. \cite{putaala2008detection} performed temperature cycling testing on ball grid array (BGA) components and monitored S-parameters. The results showed qualitative changes in S-parameters as the component degraded, while the change in the DC resistance remained negligible. Foley et al. \cite{foley2000novel} presented an approach for void detection in a transmission line by monitoring the changes in the leakage conductance parameter calculated from the S-parameter measurements. Ghaffarian et al. \cite{ghaffarian2000thermal} also conducted temperature cycling testing on the packages and characterized electrodes using S-parameters. S-parameters have also been widely used for cable fault characterization \cite{weber2020cable}. Recently, Said et al. \cite{said2022deep} proposed a deep learning-based fault classification and location for underground cables by using S-parameters. It was also shown that the cable parameters can be classified and predicted based on S-parameters using the machine learning method \cite{al2023classification}. However, none of the techniques can distinguish the causes of defects originating from various mechanisms without the aid of material characterization techniques, such as scanning electron microscopy (SEM), transmission electron microscopy (TEM), X-ray Photoelectron Spectroscopy (XPS), etc. In addition, the aforementioned studies with high-frequency parameters have utilized the parameter values at designated frequencies, not the patterns of the parameters obtained in the full range of operating frequencies. In this context, predetermined correlations between the parameters and the evolution of defects are susceptible to unexpected industrial noises. Yet, systematic studies of the effects of noise on diagnostic accuracy have not been provided.

To address the issue, this study proposes the use of S-parameters which are electrical parameters obtained in the RF domain, as a tool for fault diagnosis, allowing simultaneous root cause analysis and severity determination of defects in ITO electrodes. Valuable information regarding the severity and causes of the defect can be extracted by focusing on specific S-parameter patterns of each defective electrode. S-parameters describe the electrical behavior of networks when being subjected to various electrical signals, encompassing the electrical properties of components such as inductors, capacitors, and resistors \cite{kang2019early}. The key patterns in the S-parameter data can indicate specific features in the electrodes \cite{kang2021quantification, kang2023learning, suh2024fault}. Moreover, the S-parameter patterns provide us with two-dimensional information, such as magnitudes and frequencies, instead of single values at specific measurement times, making them suitable for machine learning (ML) techniques. If particular defect causes and severity levels exhibit specific S-parameter patterns, machine learning algorithms can be developed to detect defects in the electrodes with simultaneous identification of their root causes and severity levels. In this work, the feasibility of using S-parameter patterns is demonstrated and deep learning (DL) methods are applied to the acquired S-parameter data under various defect labels. Also, it is expected that diagnostic accuracy and noise robustness can be further improved by combining different channels, which serve as the two main pillars of S-parameters in two-port systems. 


    The major contributions of this study are as follows:
    \begin{itemize}
        \item S-parameters for various defect states of ITO electrodes are obtained and distinguishable patterns related to different defect categories are identified.
        \item By applying deep learning (DL) techniques to the acquired S-parameter patterns, we achieve high diagnostic performance and demonstrate the capability of root cause analysis.
        \item We use the S-parameter patterns obtained in a full range of the operating frequencies to provide robust diagnostic performance under noisy conditions.        
        \item Different channels are combined to enhance the diagnostic performance of the fault detection method based on the S-parameter pattern analysis.
    \end{itemize}

\section{Experimental Methodology}
	\label{sec:method}

        \subsection{Specimen Preparation}
       
\begin{figure}[!t]
		\begin{center}
			\includegraphics[width=\columnwidth]{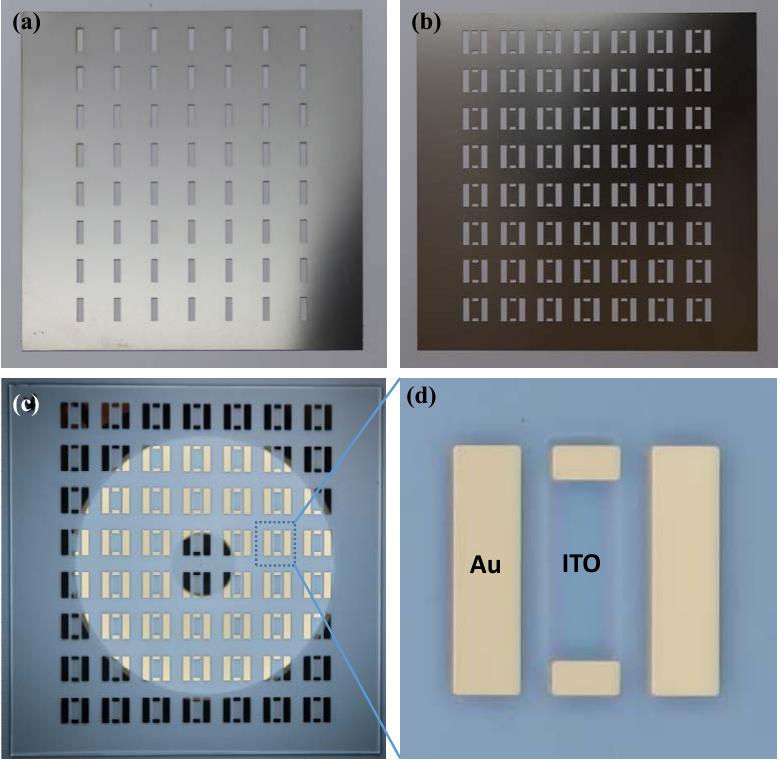}
			\caption{Fabrication of ITO electrodes and the specimen batch} 
			\label{fig:ITO}
		\end{center}
	\end{figure}

        In this study, ITO electrodes were fabricated, as shown in \cref{fig:ITO}. The fabrication procedures are as follows. (i) ITO is deposited onto the shadow mask (\cref{fig:ITO}(a)) attached to the glass substrate. (ii) The shadow mask is removed, and the other shadow mask is attached (\cref{fig:ITO}(b)) for ground lines and pads to the substrate. (iii) Chrome is deposited as a base layer, and gold is then deposited onto the shadow mask. As a result, a total of 56 specimens were made on a glass substrate, as shown in \cref{fig:ITO}(c), to obtain as much data as possible with a single batch. \cref{fig:ITO}(d) shows a single specimen. The signal and ground lines are 7 mm long and 2 mm wide. Each contact pad is 1 mm long and 2 mm wide. The gold contact pads were deposited on ITO signal lines, and gold electrodes on both sides were needed to utilize the ground--signal--ground probe tips for the S-parameter measurements. Moreover, the gold pads and electrodes enable measurements exclusively focusing on the effects of the defects in the ITO electrodes.

\begin{figure}[!t]
		\begin{center}
			\includegraphics[width=\columnwidth]{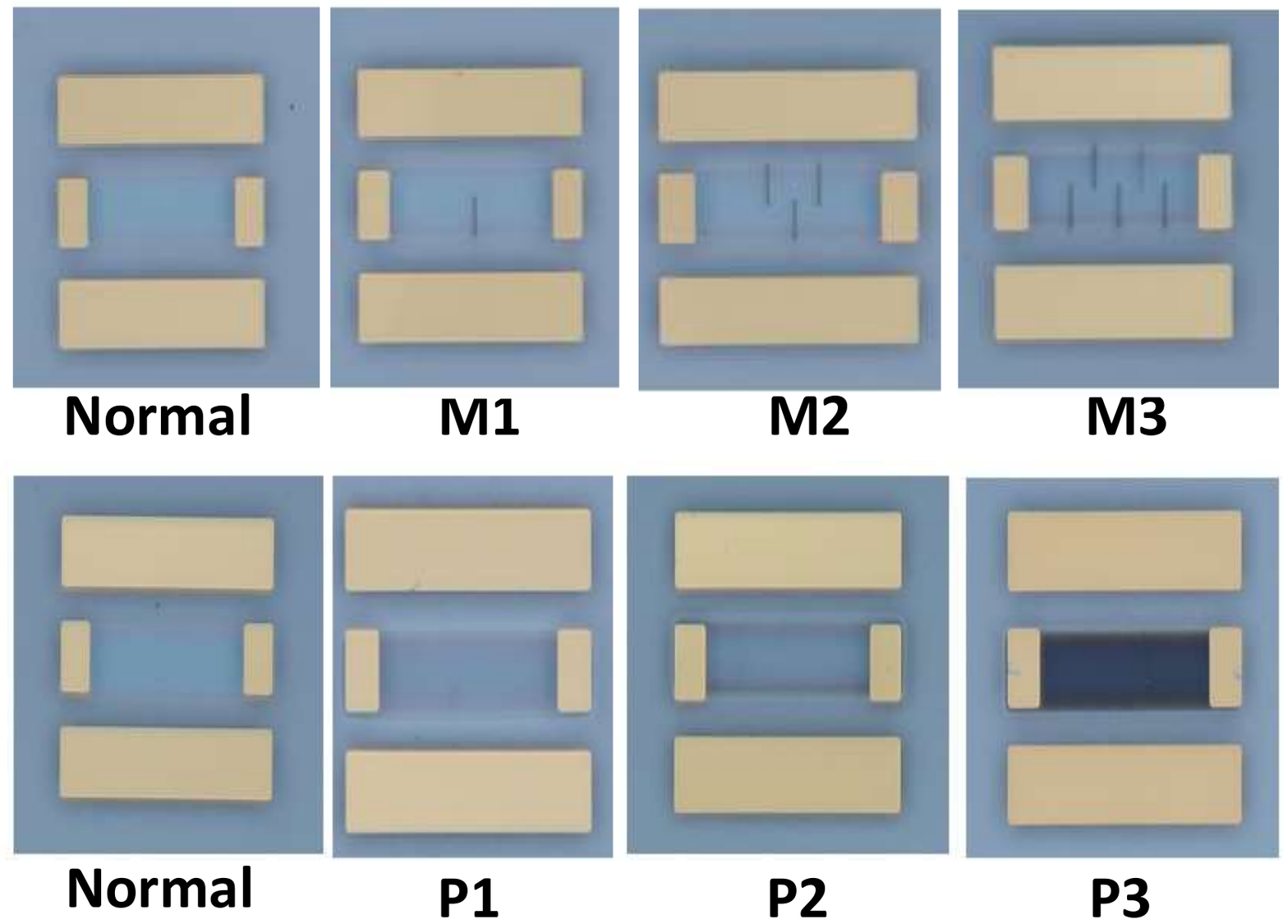}
   \vspace{-5mm}
			\caption{ITO electrodes with various defect states} 
			\label{fig:label}
		\end{center}
	\end{figure}

        Most optoelectronic part failures are related to packaging, which is susceptible to environmental stresses. For instance, external shock and vibration produce mechanical stresses that may cause material fatigue, leading to crack evolution. For the optoelectronic components, irradiation can cause the photodegradation of the packaging materials. Among the many failure modes described above, the two representative root causes of defects, namely, crack and photodegradation, are considered. \cref{fig:label} shows the ITO specimens with the induced defects and labels of the defect states. According to the level of severity, the defects were classified into four levels: the normal state, level 1 (defective but still usable), level 2 (highly recommended for replacement), and level 3 (out of order). As an example of mechanical defects, for the electrodes, 1-mm-long and 10-$\mu$m-wide cracks were precisely induced in the specimens with a laser cutting machine. The specimens with one, three, and five cracks were labeled M1, M2, and M3, respectively, representing the severity levels of the mechanical defects. Next, the specimen batch was exposed to the environmental profile according to the MIL-STD-810G solar radiation method to produce photodegradation defects in the ITO specimens. An accelerated weathering tester (QUV, Q-Lab) provided the environmental conditions. The specimens were then photographed every 100 h. The electrode specimens with 150 h, 500 h, and 1000 h were classified as P1, P2, and P3, respectively, representing the severity levels of the photodegradation defects. 
886 samples were used in total. In terms of the defect labels, 220 samples for Normal, 110 for M1, 112 for M2, 112 for M3, 110 for P1, 110 for P2, and 112 for P3 were fabricated.

        \begin{figure}[!t]
		\begin{center}
			\includegraphics[width=\columnwidth]{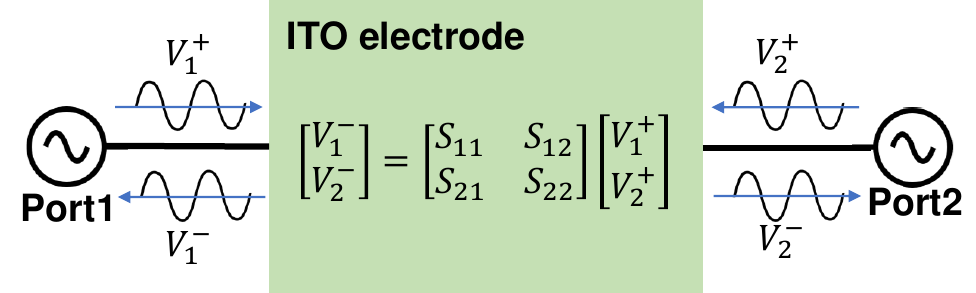}
   \vspace{-5mm}
			\caption{Two-port network of S-parameters comprised of the ITO electrode} 
			\label{fig:2port}
		\end{center}
 \vspace{-5mm}
	\end{figure}
 
        \subsection{Acquisition of S-parameters of ITO electrodes}                

         S-parameters are used as values to represent the transfer characteristics from one stage to another usually in RF electronics. A circuit, which consists of an ITO electrode and two ports for both input and output, is introduced to utilize S-parameters, as shown in \cref{fig:2port}. $V_1^+$ and $V_1^-$ and $V_2^+$ and $V_2^-$ represent the incident voltage waves and reflection on the input port (port 1) and the output port (port 2), respectively. These values can be expressed in the S-parameters of the two-port network as follows:		
 \vspace{1.5mm}
        \begin{equation}
            \begin{split}
                V_1^-&=S_{11} V_1^+ + S_{12} V_2^+,    \\
                V_2^-&=S_{21} V_1^+ + S_{22} V_2^+,     \\
                S_{11}&=(V_1^-)/(V_1^+ ) ~\text{when}~ V_2^+=0, \\
                S_{21}&=(V_2^-)/(V_1^+ ) ~\text{when}~ V_2^+=0\\
                S_{12}&=(V_1^-)/(V_2^+ ) ~\text{when}~ V_1^+=0, \\
                S_{22}&=(V_2^-)/(V_2^+ ) ~\text{when}~ V_1^+=0
            \end{split}
            \label{eq:spara}
        \end{equation}
         \vspace{1.5mm}
        
        The S-parameters consist of S11, S12, S21, and S22 channels. In symmetric systems, S21 and S12 are reciprocal (S21 = S12) and the input and output reflection coefficients are equal (S22 = S11), leaving S11 and S21 as independent channels.
        S11, also known as the return loss, indicates the signal that returned to the incident port. S21, also known as the insertion loss, shows the signal that is transmitted to the output port. The behavioral model of the electrode is buried in the stimulus response of these voltage waveforms. The electrical behaviors include resistance, capacitance, inductance, and changes in electrical properties resulting from physical damage. The S-parameter patterns are fully exploited throughout this study for reliability assessment on the ITO electrodes. 
       
\begin{figure}[!t]
		\begin{center}
			\includegraphics[width=\columnwidth]{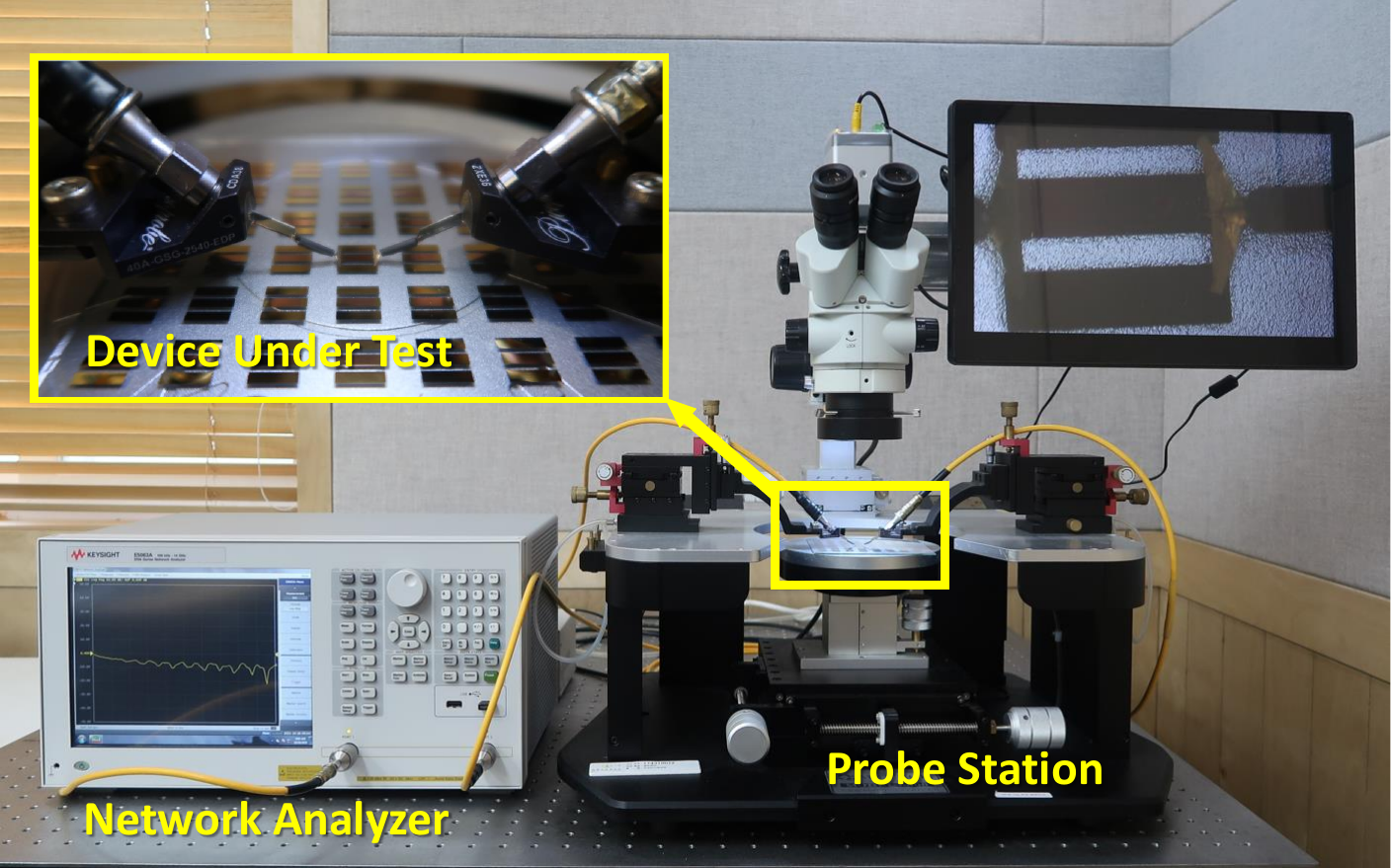}
\vspace{-7mm}
   
			\caption{Experimental setup for obtaining S-parameters} 
			\label{fig:experimentalsetup}
		\end{center}
  	\end{figure}

        \cref{fig:experimentalsetup} shows the design of an experimental setup to investigate the S-parameter indication of various defects in the ITO electrodes. The batch-type test vehicle, containing a total of 56 specimens, allowed convenient measurement using a probe station and uniformly exposed each specimen to environmental stress. The batch-type test vehicle was placed on the probe station (MSTECH Model 5500), and the S-parameters were obtained. This experimental setup helped avoid the degradation of RF cables and connectors by concentrating only on the S-parameter patterns related to the defect evolution in the test vehicle. Both ends of the specimen were contacted by high-frequency probes (ground--signal--ground type, GCB Industries Inc. 40A-GSG-2540-EDP) connected to each port of a vector network analyzer (KEYSIGHT E5063A), indicating that they formed a two-port network. The operating frequency range of the network analyzer is from 100 kHz to 18 GHz. The S11 and S21 parameter patterns were gathered according to the defect states to investigate the S-parameter indication of various causes and severity levels of defects in the ITO electrodes. In addition, the DC resistance between the ends of the specimen was measured using a digital multimeter (Fluke 1587 FC) and compared with the S-parameter measurement results.

         \begin{figure}[!t]
		\begin{center}
			\includegraphics[width=\columnwidth]{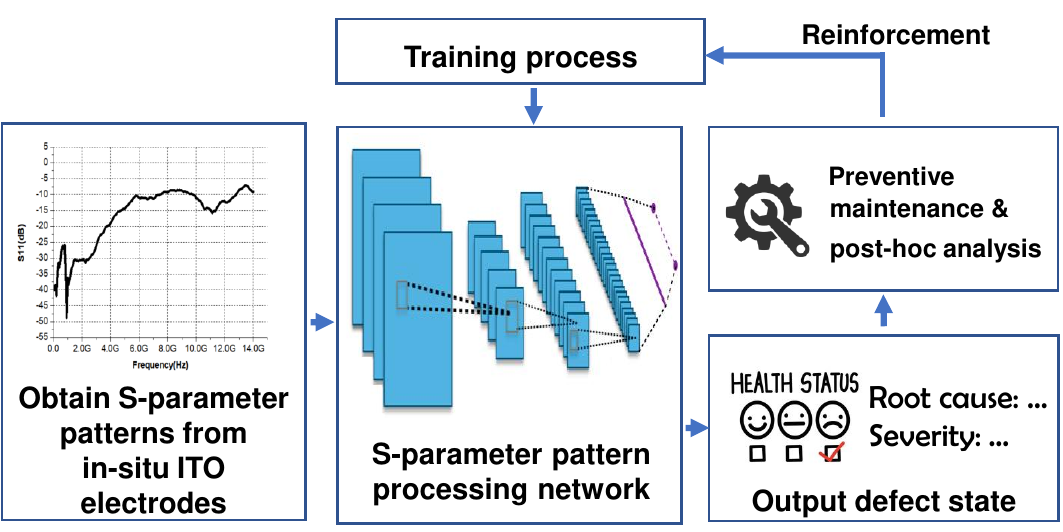}
			\caption{Process for proposed in-situ fault diagnosis and follow-up measures} 
			\label{fig:process}
		\end{center}
	\end{figure}


	\subsection{In-situ fault diagnosis using S-parameter patterns}     
         If certain patterns in the S-parameters indicate specific features of the reliability of the ITO electrodes, learning to recognize such patterns would enable observers to translate the S-parameters and then respond to the situation immediately. In that case, DL methods can be effective tools for recognizing the patterns, especially in industrial environments where measurement noises would matter. Suggested as \cref{fig:process} shows a process of using the proposed method to analyze the cause and severity of defects in the ITO electrodes in industrial applications. Firstly, to train the network, one should obtain S-parameter patterns with regard to the defect causes and severity levels. After users adjust the types of root causes and the number of severity levels to their own purpose, training data should be gathered accordingly. The users can then train and build the learning algorithm. At this point, the trained network can be deployed to industrial fields. Secondly, whenever users monitor S-parameter patterns of components of interest, they can feed the pattern to the network. The proposed method can then decide the root cause and severity level of the defects based on the trained model. The users can respond to the situation as soon as possible using the information about the defects.
         Furthermore, the root cause analysis result of the proposed method can help the users minimize the need for repeated repairs, warranty claims, and product recalls, leading to cost savings in the long term. Lastly, if the prediction is inaccurate based on a postmortem analysis, the users can reinforce the pretrained network with the relabeled S-parameter patterns. It should be noted that the entire process is based on the S-parameter patterns obtained between both ends of the ITO electrodes, enabling in situ fault diagnosis.  

    \subsection{Implementation Details}
        
       \begin{table}[t]
		\begin{center}
            \caption{The number of training and test datasets}
            \label{tab:testdatasets}
            \begin{tabular}{c|c|c|c|c}
                \hline
                Class   &   \makecell[l]{\hspace{1mm}Training\\(No noise)}   &   \makecell[l]{\hspace{4mm}Test\\(0dB noise)}   &  \makecell[l]{\hspace{4mm}Test\\(5dB noise)}  &  \makecell[l]{\hspace{5mm}Test\\(10dB noise)}  \\ \hline
                Normal  &   220        &   1080        &   1080      &  1080  \\
                M1      &   110        &   520         &   600       &  560  \\
                M2      &   112        &   560         &   560       &  560  \\
                M3      &   112        &   560         &   560       &  560  \\
                P1      &   110        &   520         &   560       &  560  \\
                P2      &   110        &   520         &   560       &  560  \\
                P3      &   112        &   560         &   560       &  560  \\ \hline
			\end{tabular}	
		\end{center}
	\end{table}
 
        \begin{figure}[!t]
		\begin{center}
			\includegraphics[width=0.9\columnwidth]{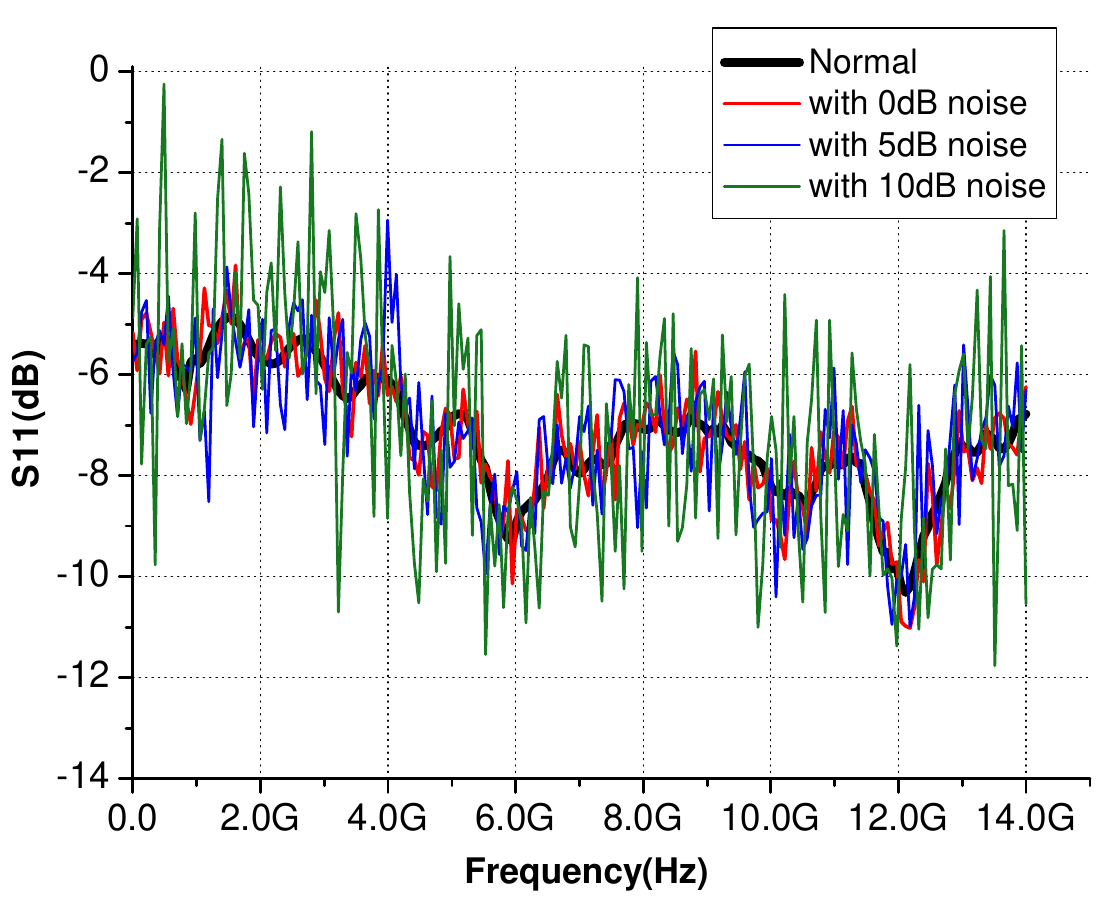}
                \vspace{-3mm}    
			\caption{Signal patterns of normal and noisy data} 
                
			\label{fig:resultsnoise}
		\end{center}
	\end{figure}
 
        In this study, three distinct DL methods, i.e., multilayer perceptron (MLP), convolutional neural network (CNN), and transformer methods, were applied to learn S-parameter patterns and were explored to classify the cause and severity of defects. First, an MLP composed of six dense layers with 512, 512, 256, 128, 64, and 7 channels, respectively, was employed. Each dense layer, except for the final one, was followed by a dropout layer with a rate of 0.2 and a Gaussian error linear unit (GELU) activation function. The final layer was applied with a softmax activation function to facilitate classification into seven possible categories. Second, a CNN approach is adopted, with the primary structure inspired by the EfficientNet architecture~\cite{tan2019efficientnet}. Beyond the default parameters, the number of layers was adjusted to eight, and the channels numbered 256, 256, 256, 256, 256, 512, 512, and 1024, respectively. Lastly, a transformer network, which contained only an encoder with 12 layers, was utilized. Each layer in the encoder consisted of four heads with 512 channels. The resulting encoded feature vector was refined via an average pooling layer and provided the classification output through a softmax activation function. The early stopping method was used to determine the number of epochs. An Nvidia RTX 3090 Ti graphics card was used to perform all experiments.
     
        \begin{figure}[!t]
		\begin{center}
            \subfloat[S11 patterns of mechanical defects]{\includegraphics[width=0.5\columnwidth]{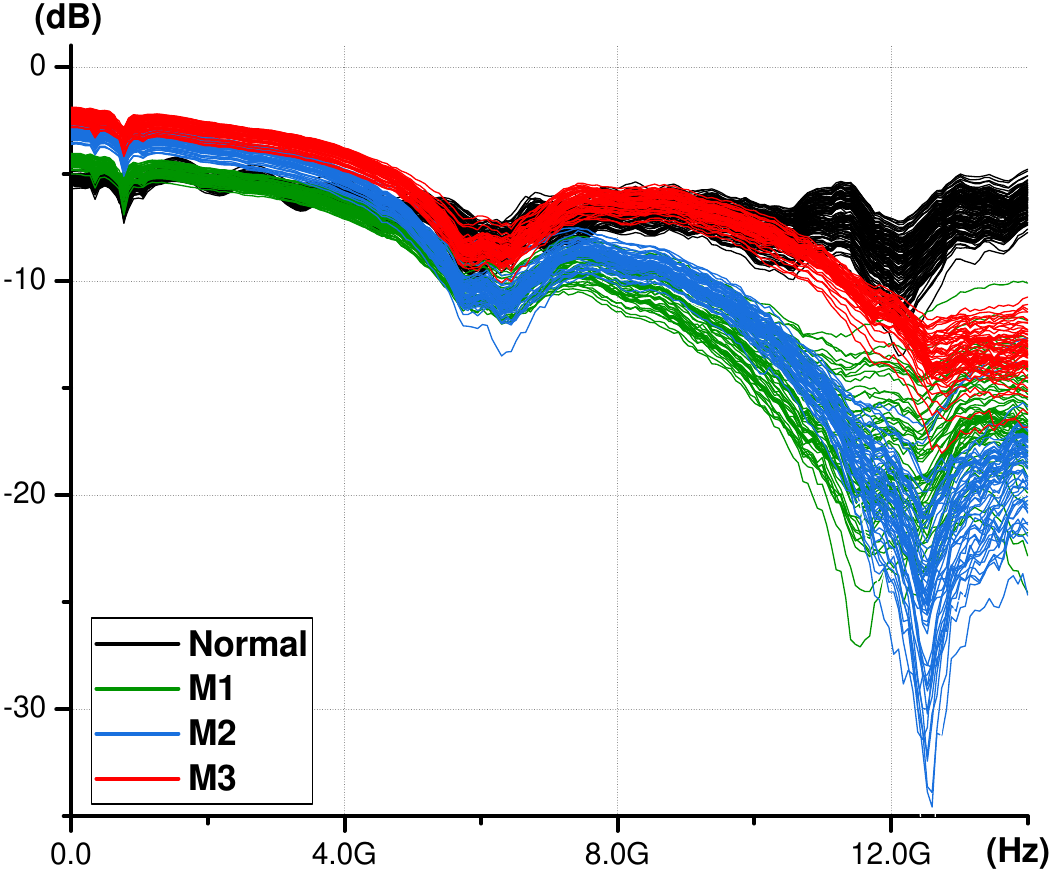}}           
            \subfloat[S11 patterns of photodegradation]{\includegraphics[width=0.5\columnwidth]{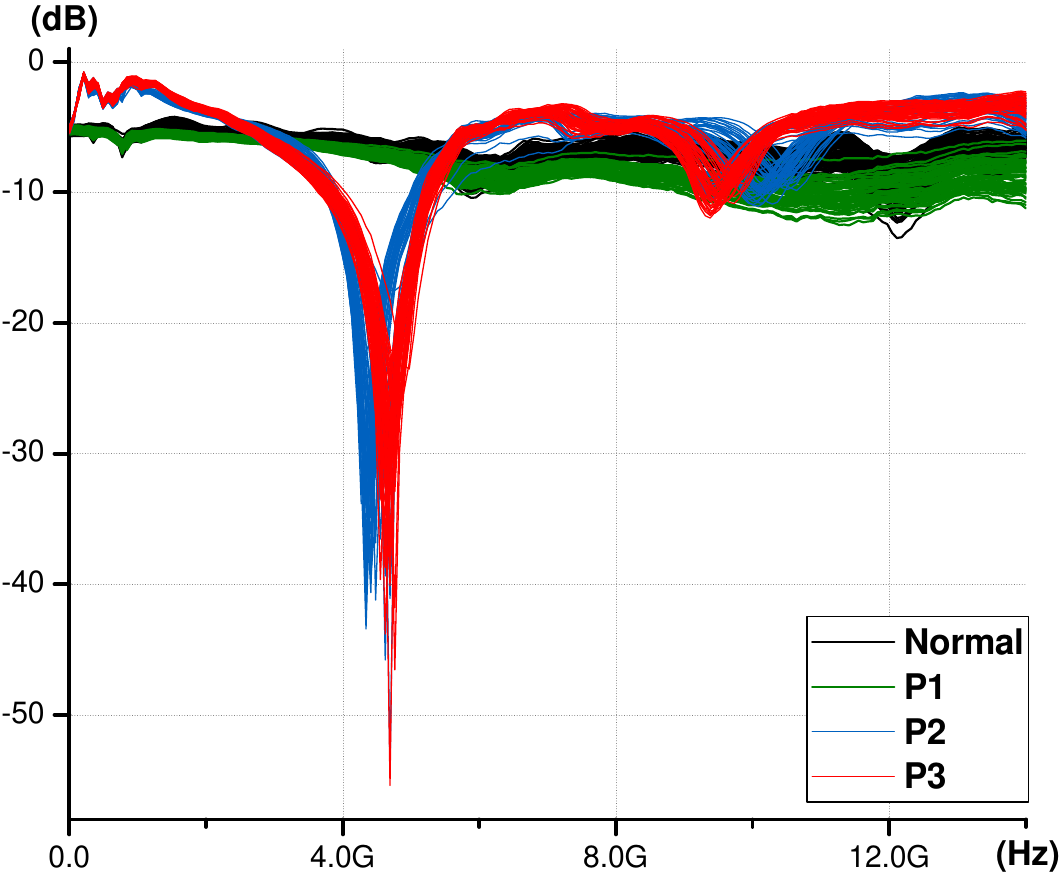}}           
            \hfill
            \subfloat[S21 patterns of mechanical defects]{\includegraphics[width=0.5\columnwidth]{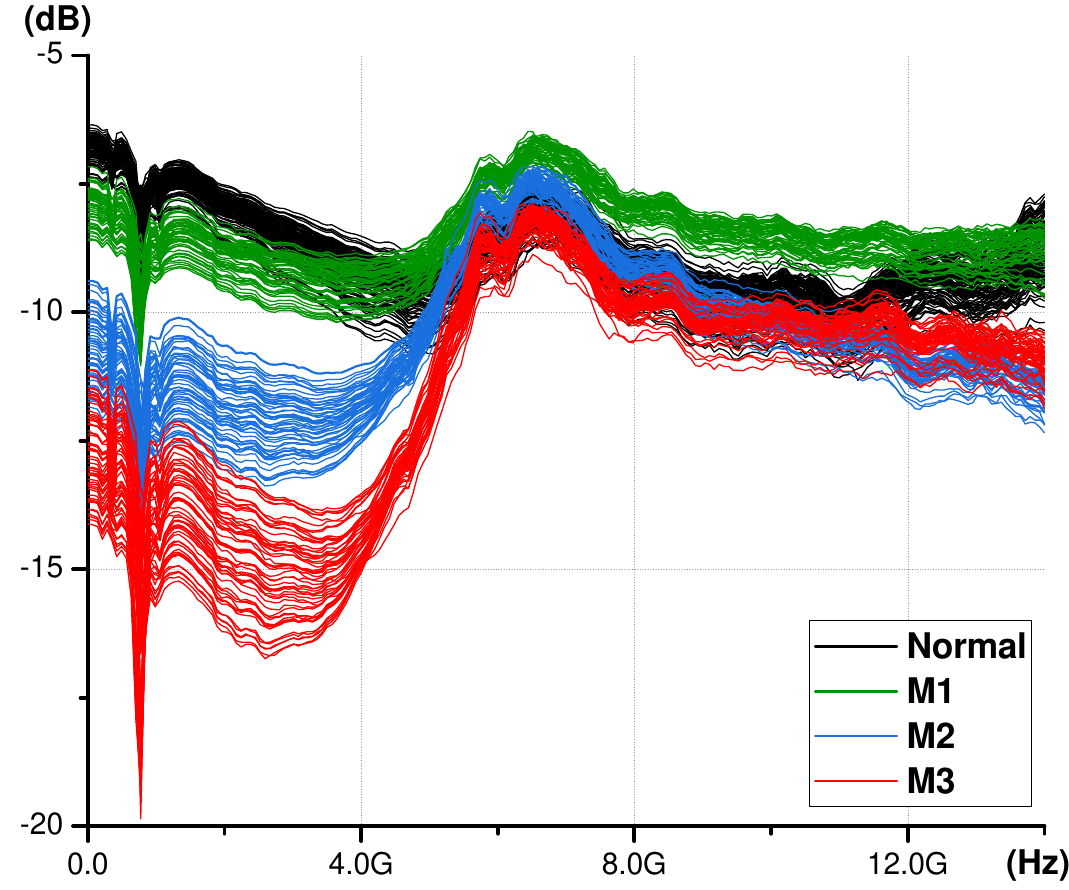}}           
            \subfloat[S21 patterns of photodegradation]{\includegraphics[width=0.5\columnwidth]{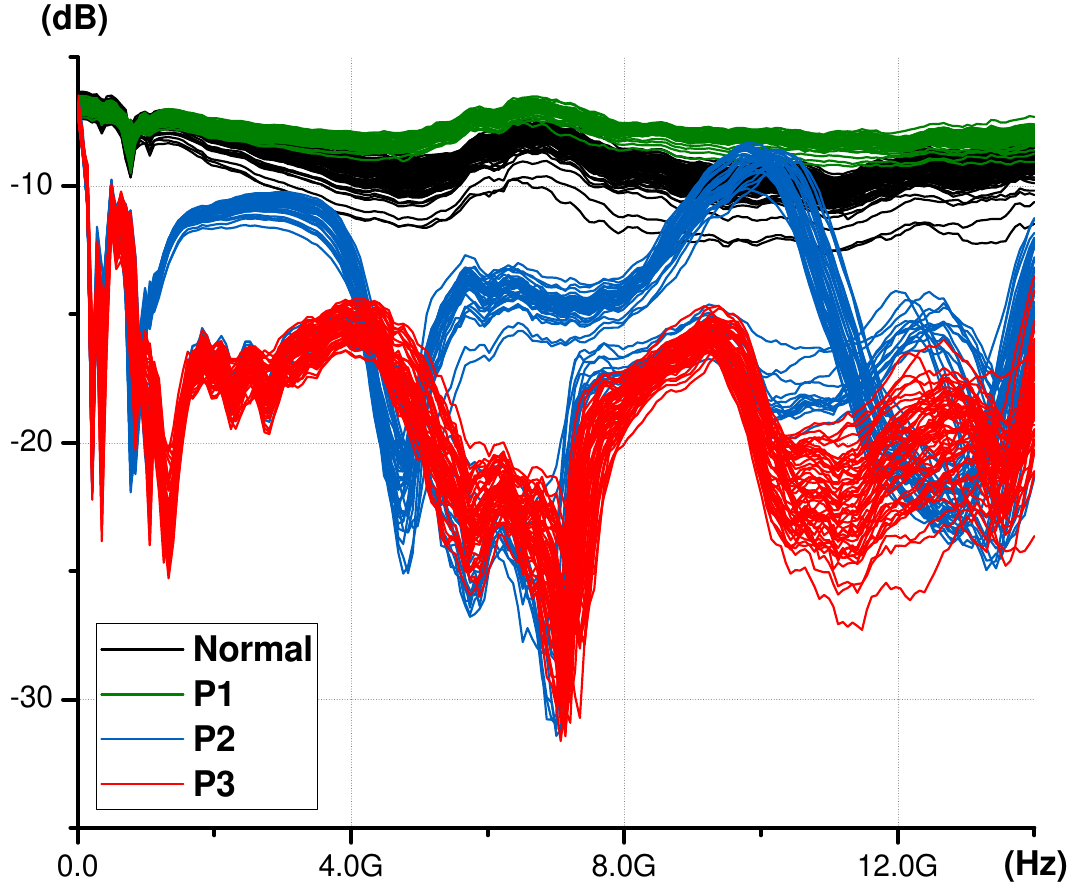}}           
            \hfill
            \caption{S-parameter patterns according to defect classes}
            \label{fig:spatterns}
		\end{center}
        \end{figure}
        \begin{figure}[!t]
		\begin{center}
			\includegraphics[width=0.75\columnwidth]{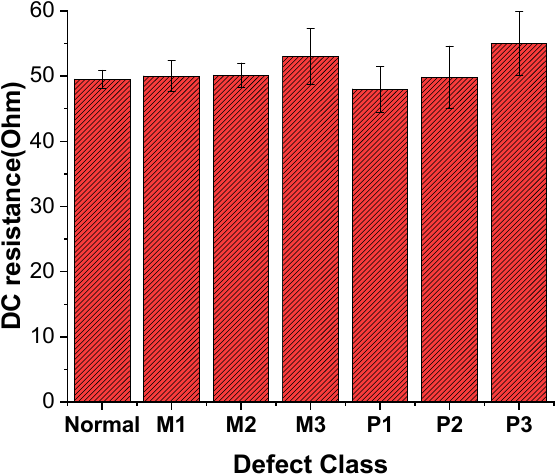}
                \vspace{-2mm}
			\caption{DC resistances according to defect classes} 
			\label{fig:dcresistances}
		\end{center}
	\end{figure}
 
        \cref{tab:testdatasets} summarizes the number of data used in training, validating, and testing. In particular, 5-fold cross-validation was conducted. The dataset comprised of 886 real samples and 4320, 4480, and 4440 noise-injected data generated by adding $0\,dB$, $5\,dB$, and $10\,dB$ noises to the real dataset, respectively. It should be noted that 0 dB does not mean zero noise but a noise level equal to the reference signal power. The compared models were trained with experimentally observed S-parameter data with no noise and were tested with noise-injected data. Various noises inevitably exist in the real world. Therefore, it is necessary to develop robust diagnosis models for various noisy data. In this study, white noise is added to data samples to reflect real-world values of electrical sensors and measuring devices. Electrical sensors have thermal, contact, and electrical noises, which are equivalent to white noise \cite{aziz2020novel}. White noise is a random signal with an equal intensity over the full spectrum. The approach used for noise generation in this study is expressed as follows \cite{30}: 
        
        \begin{equation}
            S_{noisy} = S_{observed} + \text{wgn}(x)         
        \end{equation}
        
        where $S_{noisy}$ denotes S-parameters with additional noise, $S_{observed}$ denotes the original S-parameters observed from the experiments, and the wgn(x) function is the white Gaussian noise function embedded in MATLAB, which provides Gaussian power noise with the signal power of x in decibels. \cref{fig:resultsnoise} shows the S11 parameter data of a normal ITO electrode and noisy datasets, which are the results of adding 0dB, 5dB, and 10dB white noises. Table 1 summarizes the training and test data. A 5-fold cross-validation was conducted on the 443 real samples, and the performance of the three networks was evaluated on the noisy test sets to mirror a real-world environment. Precision, recall, accuracy, and the F1 score were used to assess the diagnostic performance. The F1 score, to be specific, combines the precision and the recall using their harmonic mean, and maximizing the F1 score implies simultaneously maximizing both of them. These metrics are computed as follows: precision = TP/(TP+FP), recall = TP/(TP+FN), accuracy = (TP+TN)/(TP+FN+FP+TN), and F1 score = 2 $\times$ precision $\times$ recall / (precision + recall), where TP, FP, TN, and FN denote true positives, false positives, true negatives, and false negatives, respectively.


        \begin{table*}[!t]
		\begin{center}
            \caption{Diagnostic performances of the compared methods tested with additional noise levels (Bold numbers indicate the highest values under the same additional noise level and model. It should be noted that $0\,dB$ does not mean zero noise.)}
            \label{tab:performance}
            \renewcommand{\arraystretch}{1.3} 
            \begin{tabular}{c|c|ccc|ccc|ccc}
                \hline\hline
                \multirow{2}{*}{Model} &  \multirow{2}{*}{Metrics}&   \multicolumn{3}{c|}{S11+S21} & \multicolumn{3}{c|}{S11} & \multicolumn{3}{c}{S21} \\ \cline{3-11}
                & & 0\,dB & 5\,dB & 10\,dB & 0\,dB & 5\,dB & 10\,dB & 0\,dB & 5\,dB & 10\,dB \\ \hline
                
                \multirow{8}{*}{MLP}	& \multirow{2}{*}{Precision} & \textbf{93.27} & \textbf{74.37} & \textbf{37.76} & 83.53 & 71.45 & 29.78 & 77.65 & 53.30 & 22.01 \\ 
                ~ & ~ & $\pm1.93$ & $\pm4.85$ & $\pm2.44$ & $\pm3.80$ & $\pm4.99$ & $\pm1.31$ & $\pm2.54$ & $\pm2.27$ & $\pm1.98$ \\ \cline{2-11}
                & \multirow{2}{*}{Recall} & \textbf{93.26} & \textbf{74.05} & 30.03 & 83.81 & 68.39 & \textbf{31.64} & 73.65 & 52.05 & 24.41 \\ 
                && $\pm2.07$ & $\pm4.65$ & $\pm3.56$ & $\pm3.77$ & $\pm5.30$ & $\pm1.93$ & $\pm1.98$ & $\pm2.24$ & $\pm0.57$ \\ \cline{2-11}
                & \multirow{2}{*}{Accuracy} & \textbf{93.34} & \textbf{72.43} & 27.90 & 85.88 & 63.24 & \textbf{28.69} & 70.86 & 47.84 & 21.87 \\ 
                && $\pm1.75$ & $\pm3.74$ & $\pm3.23$ & $\pm3.11$ & $\pm4.03$ & $\pm2.10$ & $\pm2.19$ & $\pm2.45$ & $\pm0.59$\\ \cline{2-11}
                & \multirow{2}{*}{F1} & \textbf{92.96} & \textbf{72.51} & \textbf{28.00} & 82.68 & 64.82 & 25.69 & 70.89 & 42.88 & 15.45  \\ 
                && $\pm1.90$ & $\pm4.36$ & $\pm3.48$ & $\pm4.28$ & $\pm5.25$ & $\pm2.37$ & $\pm2.89$ & $\pm3.18$ & $\pm1.07$\\\hline
                
                \multirow{8}{*}{CNN}	& \multirow{2}{*}{Precision} & \textbf{97.34} & \textbf{99.18} & \textbf{90.70} & 91.46 & 92.77 & 74.00 & 90.08 & 78.55 & 50.07 \\ 
                && $\pm1.56$ & $\pm0.33$ & $\pm4.64$ & $\pm3.80$ & $\pm1.17$ & $\pm3.04$ & $\pm1.36$ & $\pm5.39$ & $\pm11.94$\\ \cline{2-11}
                & \multirow{2}{*}{Recall} & \textbf{96.74} & \textbf{99.06} & \textbf{90.89} & 91.14 & 93.01 & 62.11 & 89.24 & 72.65 & 39.69 \\ 
                && $\pm1.96$ & $\pm0.36$ & $\pm6.62$ & $\pm3.70$ & $\pm1.16$ & $\pm4.35$ & $\pm1.73$ & $\pm8.60$ & $\pm9.02$ \\ \cline{2-11}
                & \multirow{2}{*}{Accuracy} & \textbf{97.13} & \textbf{99.12} & \textbf{87.58} & 92.21 & 93.21 & 57.69 & 88.77 & 70.56 & 35.10 \\ 
                && $\pm1.73$ & $\pm0.32$ & $\pm9.80$ & $\pm3.03$ & $\pm1.07$ & $\pm4.94$ & $\pm2.68$ & $\pm9.08$ & $\pm8.03$\\ \cline{2-11}
                & \multirow{2}{*}{F1} & \textbf{96.88} & \textbf{99.11} & \textbf{88.83} & 90.72 & 92.73 & 58.13 & 88.66 & 69.03 & 30.36  \\ 
                && $\pm1.88$ & $\pm0.34$ & $\pm8.88$ & $\pm4.42$ & $\pm1.18$ & $\pm4.70$ & $\pm2.03$ & $\pm10.14$ & $\pm6.78$\\\hline
                
                \multirow{8}{*}{Transformer}	& \multirow{2}{*}{Precision} & \textbf{92.13} & \textbf{96.50} & \textbf{84.60} & 89.83 & 94.07 & 80.51 & 76.98 & 83.17 & 60.34 \\ 
                && $\pm8.65$ & $\pm1.10$ & $\pm1.14$ & $\pm4.82$ & $\pm0.92$ & $\pm1.47$ & $\pm7.58$ & $\pm1.63$ & $\pm1.58$\\ \cline{2-11}
                & \multirow{2}{*}{Recall} & \textbf{92.74} & \textbf{96.91} & \textbf{85.93} & 90.03 & 94.32 & 79.27 & 80.57 & 81.81 & 64.29 \\ 
                && $\pm5.71$ & $\pm1.11$ & $\pm1.05$ & $\pm3.57$ & $\pm0.78$ & $\pm2.08$ & $\pm4.83$ & $\pm1.92$ & $\pm0.88$  \\ \cline{2-11}
                & \multirow{2}{*}{Accuracy} & \textbf{93.69} & \textbf{96.59} & \textbf{83.15} & 91.29 & 94.37 & 78.59 & 83.21 & 83.60 & 62.89  \\ 
                && $\pm4.78$ & $\pm1.16$ & $\pm2.16$ & $\pm2.93$ & $\pm0.90$ & $\pm2.20$ & $\pm4.08$ & $\pm1.73$ & $\pm1.28$\\ \cline{2-11}
                & \multirow{2}{*}{F1} & \textbf{92.17} & \textbf{96.57} & \textbf{84.20} & 89.41 & 94.08 & 77.13 & 78.07 & 81.28 & 59.08 \\ 
                && $\pm7.40$ & $\pm1.19$ & $\pm1.61$ & $\pm4.73$ & $\pm0.88$ & $\pm3.04$ & $\pm6.45$ & $\pm2.01$ & $\pm1.52$\\\hline\hline
			\end{tabular}
		\end{center}
	\end{table*}          
 
\section{Experimental results}







        \subsection{Signal measurement results}
         \cref{fig:spatterns} shows the S-parameter patterns according to the defect states. In comparison, \cref{fig:dcresistances} depicts the associated values of DC resistances of the electrodes according to the occurrence and evolution of the defects. In terms of mechanical defects, as the number of cracks in the ITO electrodes increased from 0 to 5, i.e., from normal to M3 states, the S-parameters showed distinguishable patterns for the mechanical defects. Similarly, as the degree of photodegradation aggravated from normal, P1, P2, to P3 classes, the S-parameters of photodegraded ITO electrodes also showed unique patterns for the type of defects. Meanwhile, the change in DC resistance according to the defect classes was negligible. The average DC resistances of normal, M1, M2, and M3 states were 49.5, 50.0, 50.1, and 53.2$\,\Omega$, respectively. Although increasing, the change remained within the standard deviation. Therefore, it can be challenging to detect the mechanical defect and track its evolution with the DC resistance measurement. The changes in DC resistance were not prominent as the photodegradation defects progressed. The average DC resistances of normal, P1, P2, and P3 classes were 48.0, 49.8, and 55.1 $\,\Omega$, respectively. Again, although this seemed to increase, the change remained within the standard error. Overall, defective ITO electrodes exhibited unique S-parameter patterns according to the characteristics of defects. The electrodes with the same cause of the defects showed similar patterns, and the severity level induced minor changes in the patterns, such as magnitude offsets and peak shifts. The measurement results on S-parameter patterns and DC resistances show that the former provides the capabilities of early detection and root cause analysis, while the latter is insensitive to the causes of the defects.

        \subsection{Quantitative Evaluation on Diagnostic Performance}

        \begin{figure*}[!t]
		\begin{center}
         \includegraphics
           [width=1.0\linewidth]
           {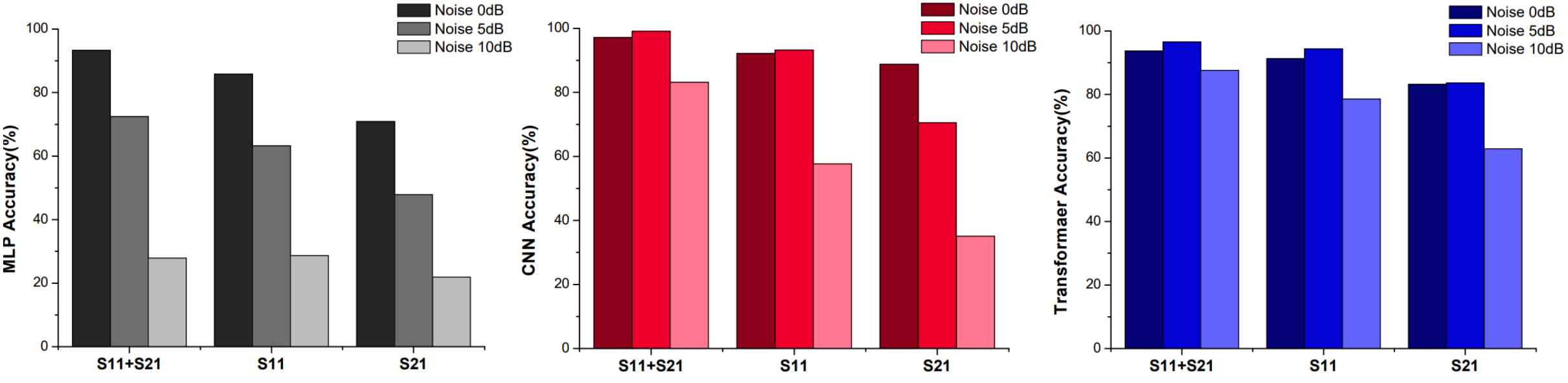}       
             
			\caption{Diagnostic performance of MLP, CNN, and Transformer models with different combinations of S-parameters under additive noise} 
			\label{fig:accuracy}
		\end{center}
	\end{figure*}

        \cref{tab:performance} and \cref{fig:accuracy} summarize the diagnostic performances of the compared models under various additional noise levels and combinations of S-parameter channels. Precision, recall, accuracy, and the F1 score are used as metrics to evaluate the diagnostic performances of the compared models. These values are expressed in percentages and displayed as a mean value $\pm$ standard deviation. Bold numbers represent the highest values among various combinations of S-parameter channels for the given model, metrics, and noise level. Overall, the trends across all four metrics are consistent, indicating no significant skewness effect. All the compared models provided 99.9\% of diagnostic accuracy in the training phase with observed S-parameters with no additional noise. The models were then evaluated from two perspectives, namely, the impact of additional noise levels and the influence of different S-parameter channels. The main goal for developing effective fault diagnosis techniques is to enhance the reliability of electronics in real-world industrial applications~\cite{kim2020direct}. One challenge in these settings is the random environmental and operational noise. To mimic these conditions, white noise was added to the data samples, reflecting the values typically observed in electrical sensors and measuring devices~\cite{aziz2020novel}. The results indicate that increased noise levels reduce diagnostic accuracy due to variations in test data characteristics and distributions.

        \begin{figure}[!t]
		\begin{center}
			\includegraphics[width=0.75\columnwidth]{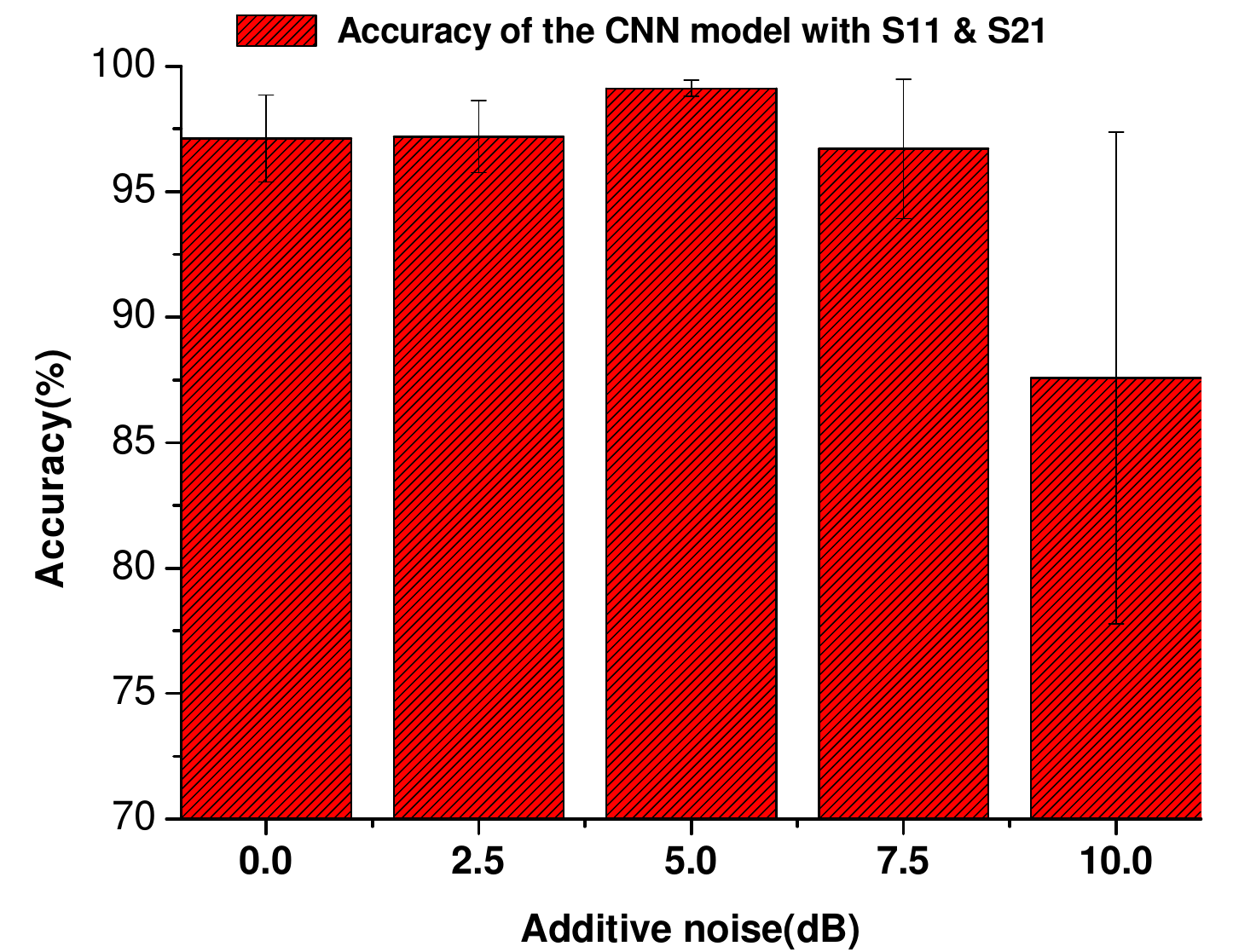}
                \vspace{-2mm}
			\caption{Effect of additive noise levels on the accuracy of the CNN model with S11 \& S21 data} 
			\label{fig:noise effect}
		\end{center}
	\end{figure}
 
        The CNN model achieved the best diagnostic performance, reaching $99.12\,\%$ accuracy under 5\,dB of additive noise using a combination of S11 and S21 parameters. Generally, the CNN model consistently displayed the highest diagnostic performance. The MLP model exhibited a decreased accuracy as additional noise levels increased. However, the CNN and transformer models achieved optimal diagnostic performance at 5\,dB of additional noise. Using the CNN and transformer models is recommended to improve the classification performance and enhance the signal robustness against noise. In this study, an interesting trend is observed, wherein the performance of the proposed model, which was trained exclusively on noiseless data, proved to be better on test data with added 5\,dB noise as compared to the data with 0\,dB noise. This intriguing phenomenon might be attributable to certain inherent qualities of the data and the robustness of the model to slight perturbations. 
        As shown in \cref{fig:noise effect}, we performed a thorough analysis regarding the effect of the additive noise on the accuracy of the CNN model with S11 and S21 data that provided the best diagnostic performance in this study. Both the accuracy of the model and the standard deviation of the performance were optimized at the additive noise level of 5dB. The beneficial role of noise injection in learning is a consolidated concept in the field of artificial neural networks. Certain structure of noise can improve performance of the neural networks \cite{bonnasse2022categorical, semenova2022understanding, benedetti2023training, tomasini2023deep}. In this study, the benefit of the additional noise is optimized at 5\,dB, augmenting the inherent variance. The addition of higher levels of noise only degraded the performance. It is worth noting that these results were achieved under the specific configurations and test conditions of the proposed model, emphasizing the need for further exploration to fully comprehend the complex interplay between noise level and model performance on unseen data.
        
        In terms of combining different S-parameter channels, the combination of S11 and S21 parameters yielded the best diagnostic performance across all DL models in the presence of additive noise. Comparatively, the S11 parameter typically provided better classification accuracy compared with the S21 parameter. Given that S11 represents return loss -- indicating how much of the signal is reflected from a defect -- it is generally more sensitive to defect evolution than S21 which signifies insertion loss.

        \begin{figure*}[!t]
		\begin{center}
          \includegraphics[width=0.7\linewidth]{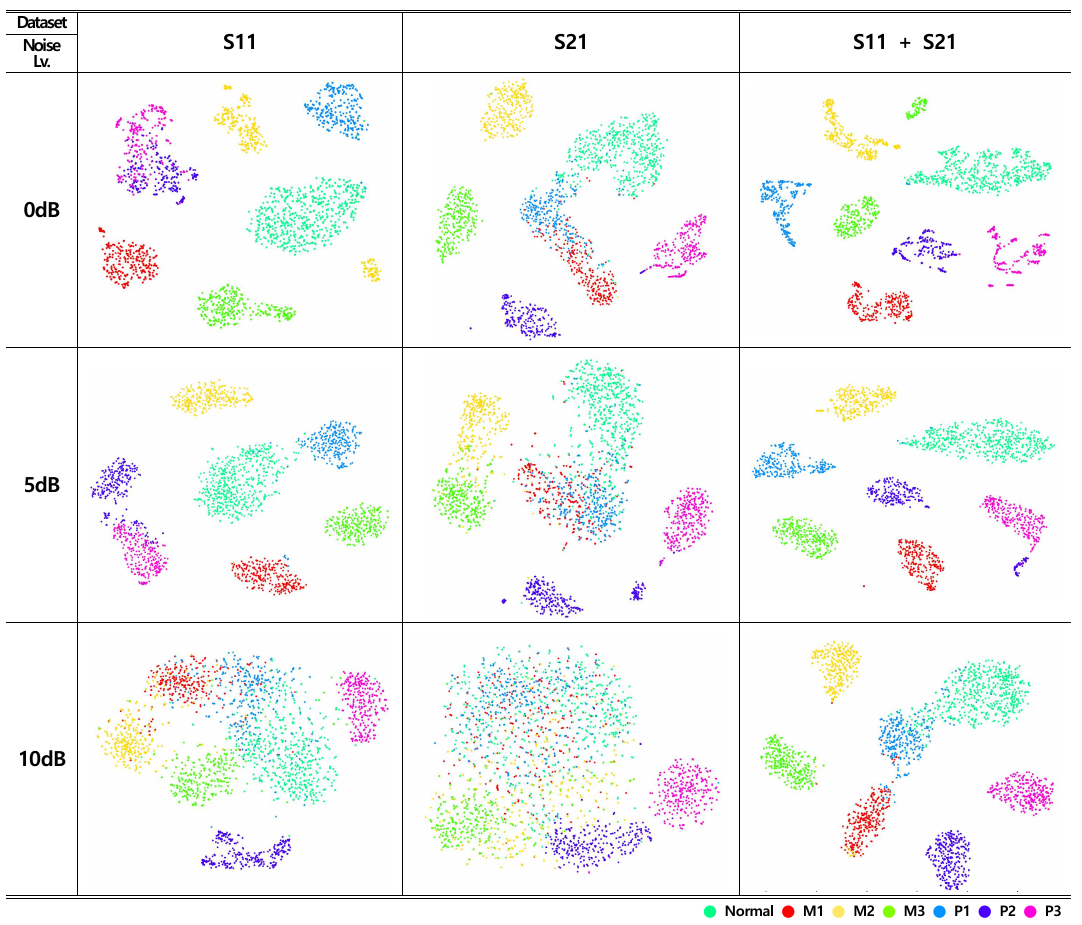}
                \vspace{-2mm}
			\caption{t-SNE dimension reduction visualization of learned features (latent vectors from the CNN model) of S-parameters with additional noise levels} 
        \end{center}
			\label{fig:tsne}
	\end{figure*}
 
\subsection{Qualitative Evaluation on Diagnostic Performance}
	\label{sec:Discussion}   
          Learned features from the S-parameters with injected noises were extracted from the CNN model as forms of latent vectors. Then the latent vectors were visualized after dimension reduction, as shown in \cref{fig:tsne}, to examine and understand the feasibility of the S-parameter patterns as inputs to the DL methods qualitatively. The dimension reduction was conducted based on the t-SNE method that maps high-dimension data into the low-dimensional embedded space \cite{van2008visualizing}. When it is difficult to interpret the meaning of the original high dimensional data, the t-SNE can be a highly effective algorithm to explain the effectiveness of the data for the learning algorithms\cite{kim2020direct}. In that, the features extracted from the S-parameter patterns were mapped into the lower-dimensional spaces to visually judge the validity of the S-parameter patterns for the fault diagnosis. 
          
          The t-SNE visualization with S-parameters with 0\,dB noise verifies the efficacy of the S-parameter patterns for diagnosing the cause and severity of defects. Specifically, the S11 channel shows more definitive boundaries of the clusters according to the defect states than the S21 channel. This indicates the superiority of the S11 channel as a precursor of the defects. The features from the combination of S11 and S21 channels result in the best clustering characteristics. This means that data representing the same health condition are well-grouped in the reduced feature space and well-separated from the data representing the other conditions. Therefore, based on the proposed method, high diagnostic accuracy could be obtained by learning the features that allow data to cluster distinctly for each health condition. However, the cluster boundaries become unclear as additive noise levels increase, implicitly indicating decreasing diagnostic performance with the increasing noise level. Especially with S21 data, M1 and P1 defects (early-stage defects) show no noticeable difference in data projection at all noise levels. Even under noisy conditions, combining S11 and S21 channels can boost classification accuracy, as can be confirmed by the t-SNE visualization trends at all noise levels.       
         
	\section{Conclusion}
	\label{sec:conclusion}   
       
        This paper proposes a novel and high-performance 
 in situ fault diagnosis model based on electrical signal patterns, specifically the S-parameter patterns. This study considers seven health states, i.e., normal, M1, M2, M3, P1, P2, and P3. Unlike traditional methods that are heavily reliant on expertise, prior knowledge, and costly material characterization, the proposed method effectively learns the relationship between fault causes and severity, enabling in situ fault detection without the need for additional tools or destructive testing.
        
        It is demonstrated that S-parameters of transparent ITO electrodes exhibit distinct patterns according to different defect causes and severity levels. The focus on photodegradation and mechanical defects in ITO electrodes reveals the capability of early detection, with negligible changes in DC resistances. The S-parameter patterns prove to be effective features for DL methods, including MLP, CNN, and transformer, achieving impressive diagnostic accuracies of 99.9\,\% without any additional noise. Even under harsher conditions with 10\,dB of noise, the CNN and transformer models show commendable noise robustness with diagnostic accuracies of 87.58\,\%, and 83.15\,\%, respectively. A significant advantage of the proposed method is its ability to simultaneously identify the causes and assess the severity of defects, avoiding the need for secondary tools for root cause analysis. This flexibility is particularly valuable in quality-critical industries, where product reliability and safety are paramount.To enhance the applicability of the proposed method in the industries, our further work will focus on developing online fault detection and diagnosis methods for ITO electrodes integrated into real optoelectronic devices, addressing measurement method challenges for obtaining S-parameter signals from these devices. The proposed method has the potential to revolutionize fault diagnosis in optoelectronic applications, enhancing product reliability and performance in practical settings.

\bibliographystyle{IEEEtran}
\bibliography{mybibfile}


    \begin{IEEEbiography}[{\includegraphics[width=1in, height=1.25in, clip,keepaspectratio]{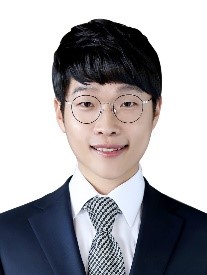}}]{Tae Yeob Kang}
		received B.S., M.S., and Ph. D. degrees in Mechanical Engineering from the Korea Advanced Institute of Science and Technology (KAIST), Daejeon, South Korea, in 2009, 2011, and 2022 respectively. He is currently an Assistant Professor at School of Industrial and Mechanical Engineering with the University of Suwon, Hwaseong, South Korea. Before joining the faculty, he had worked at Agency for Defense Development, South Korea for 12 years. His current research interests include reliability testing on defense systems, AI-assisted reliability assessment, and fault detection and diagnosis for electronic packaging.
    \end{IEEEbiography}
    \begin{IEEEbiography}[{\includegraphics[width=1in, height=1.25in, clip,keepaspectratio]{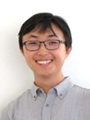}}]{Haebom Lee}
		is a doctoral student at the Heidelberg University in Germany. He received B.S. and M.S. degrees in computer science from the Korea Advanced Institute of Science and Technology (KAIST), Daejeon, South Korea, in 2009 and 2015, respectively. His research interests span deep learning applications, computer vision, synthetic data, and generative models. 
    \end{IEEEbiography}
    
    \begin{IEEEbiography}[{\includegraphics[width=1in, height=1.25in, clip,keepaspectratio]{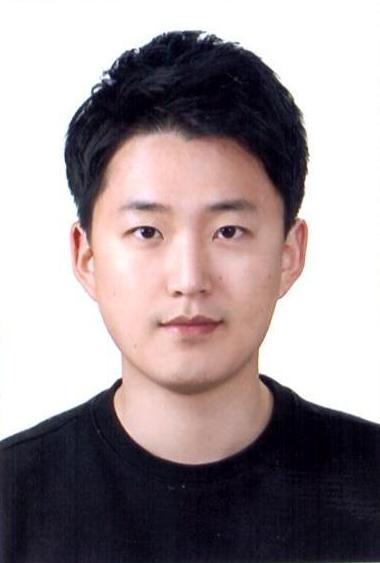}}]{Sungho Suh}
		is a Senior Researcher at the German Research Center for Artificial Intelligence (DFKI) in Germany since 2020. He received the Ph.D. degree in Computer Science at the Technische Universit{\"a}t Kaiserslautern, Germany in 2021, and the B.S. and M.S. degrees from the School of Electrical Engineering and Computer Science, Seoul National University, Seoul, South Korea, in 2009 and 2011, respectively. Before joining DFKI, he has worked at KIST Europe in Germany for three years, and at Samsung Electro-Mechanics, Korea from 2011 to 2018. His research interests are machine learning algorithms, such as sensor data processing, computer vision, multimodal processing, and generative model, with a focus on industrial applications.
    \end{IEEEbiography}

\vfill

\end{document}